\documentclass{article}
\usepackage{nips01e,epsfig}

\title{Gaussian Process Regression with Mismatched Models}

\author{Peter Sollich
\\
Department of Mathematics, King's College London\\
Strand, London WC2R 2LS, U.K. Email {\tt peter.sollich@kcl.ac.uk}
}

\newcommand{\teach}{\theta_*}
\newcommand{\stud}{\theta}
\newcommand{\meanstud}{\hat{\theta}}
\newcommand{\noise}{\sigma^2}
\newcommand{\noiseinv}{\sigma^{-2}}
\newcommand{\noisest}{\sigma_*^2}

\newcommand{\KK}{{\mathbf K}}
\newcommand{\MM}{{\mathbf M}}
\newcommand{\kk}{{\mathbf k}}
\newcommand{\yy}{{\mathbf y}}
\newcommand{\T}{^{\mathrm T}}
\newcommand{\tr}{{\rm tr\,}}
\newcommand{\eps}{\epsilon}

\newcommand{\lam}{\lambda}
\newcommand{\Lam}{\Lambda}
\newcommand{\xl}{x^l}
\newcommand{\yl}{y^l}
\newcommand{\xm}{x^m}
\newcommand{\ym}{y^m}
\newcommand{\LL}{{\mathbf \Lambda}}
\newcommand{\Ph}{{\Phi}}
\newcommand{\mident}{{\mathbf I}}
\newcommand{\GG}{{\mathbf G}}
\newcommand{\Gfluc}{{\mathcal G}}
\newcommand{\new}{{\bf \psi}}

\newcommand{\XX}{_{\{\xl\}}}

\newcommand{\ii}{s}
\newcommand{\cc}{f}

\newcommand{\bi}{\begin{itemize}\setlength{\itemsep}{0pt}}
\newcommand{\ei}{\end{itemize}}
\newcommand{\be}{\begin{equation}}
\newcommand{\ee}{\end{equation}}
\newcommand{\bea}{\begin{eqnarray}}
\newcommand{\eea}{\end{eqnarray}}
\def\(#1){(\ref{#1})}

\newcommand{\lav}{\langle}
\newcommand{\rav}{\rangle}

\newcommand{\eq}[1]{~(\ref{#1})}
\newcommand{\eqq}[2]{~(\ref{#1},\ref{#2})}

\newcommand{\order}{{{\mathcal O}}}
\newcommand{\beastar}{\begin{eqnarray*}}
\newcommand{\eeastar}{\end{eqnarray*}}

\newcommand{\ie}{{i.e.}}
\newcommand{\eg}{{e.g.}}

\newcommand{\colvec}[2]{(\!\!\!\renewcommand{\arraystretch}{0.5}\begin{array}{c}\scriptstyle#1\\\scriptstyle#2\end{array}\!\!\!)}

\begin{document}

\maketitle

\begin{abstract}

Learning curves for Gaussian process regression are well understood
when the `student' model happens to match the `teacher' (true data
generation process). I derive approximations to the learning curves
for the more generic case of {\em mismatched models}, and find very
rich behaviour: For large input space dimensionality, where the
results become exact, there are universal (student-independent)
plateaux in the learning curve, with transitions in between that can
exhibit arbitrarily many over-fitting maxima. In lower dimensions,
plateaux also appear, and the asymptotic decay of the learning curve
becomes strongly student-dependent. All predictions are confirmed by
simulations.

\end{abstract}

\section{Introduction}

There has in the last few years been a good deal of excitement about
the use of Gaussian processes (GPs) as an alternative to feedforward
networks~\cite{general_refs}. GPs make prior assumptions about the
problem to be learned very transparent, and even though they are
non-parametric models, inference---at least in the case of regression
considered below---is relatively straightforward.  One crucial
question for applications is then how `fast' GPs learn, \ie, how many
training examples are needed to achieve a certain level of
generalization performance. The typical (as opposed to worst case)
behaviour is captured in the {\em learning curve}, which gives the
average generalization error $\eps$ as a function of the number of
training examples $n$. Good bounds and approximations for $\eps(n)$
are now available~\cite{general_refs,MicWah81,Sollich99,WilViv00},
but these are all restricted to the case where the `student' model
exactly matches the true `teacher' generating the data. In practice,
such a match is unlikely, and so it is important to understand how GPs
learn if there is some model mismatch. This is the aim of this paper.

In its simplest form, the regression problem is this: We are trying to
learn a function $\teach$ which maps inputs $x$ (real-valued vectors)
to (real-valued scalar) outputs $\teach(x)$. We are given a set of
training data $D$, consisting of $n$ input-output pairs $(\xl,\yl)$;
the training outputs $\yl$ may differ from the `clean' teacher outputs
$\teach(\xl)$ due to corruption by noise. Given a test input $x$, we
are then asked to come up with a prediction $\meanstud(x)$, plus error
bar, for the corresponding output $\stud(x)$. In a Bayesian setting,
we do this by specifying a prior $P(\stud)$ over hypothesis functions,
and a likelihood $P(D|\stud)$ with which each $\stud$ could have
generated the training data; from this we deduce the posterior
distribution $P(\stud|D)\propto P(D|\stud)P(\stud)$. For a GP, the
prior is defined directly over input-output functions $\stud$; this is
simpler than for a Bayesian feedforward net since no weights are
involved which would have to be integrated out. Any $\stud$ is
uniquely determined by its output values $\stud(x)$ for all $x$ from
the input domain, and for a GP, these are assumed to have a joint
Gaussian distribution (hence the name). If we set the means to zero
(as is commonly done), this distribution is fully specified by the
{\em covariance function} $\lav\stud(x)\stud(x')\rav_\stud =
C(x,x')$. The latter transparently encodes prior assumptions
about the function to be learned.  Smoothness, for example, is
controlled by the behaviour of $C(x,x')$ for $x'\to x$: The
Ornstein-Uhlenbeck (OU) covariance function $C(x,x') =
\exp(-|x-x'|/l)$ produces very rough (non-differentiable) functions,
while functions sampled from the radial basis function (RBF) prior
with $C(x,x') = \exp[-|x-x'|^2/(2l^2)]$ are infinitely
differentiable. Here $l$ is a lengthscale parameter, corresponding
directly to the distance in input space over which we
expect significant variation in the function values.

There are good reviews on how inference with GPs
works~\cite{general_refs,Williams98}, so I only give a brief summary
here. The student assumes that outputs $y$ are generated from the
`clean' values of a hypothesis function $\stud(x)$ by adding Gaussian
noise of $x$-independent variance $\noise$. The joint distribution of
a set of training outputs $\{\yl\}$ and the function values $\stud(x)$
is then also Gaussian, with covariances given (under the student
model) by
\[
\lav \yl \ym \rav = C(\xl,\xm)+\noise\delta_{lm} = (\KK)_{lm}, \qquad
\lav \yl \stud(x) \rav = C(\xl,x) = (\kk(x))_l
\]
Here I have defined an $n \times n$ matrix $\KK$ and an $x$-dependent
$n$-component vector $\kk(x)$. The posterior distribution $P(\stud|D)$
is then obtained by conditioning on the $\{\yl\}$; it is again
Gaussian and has mean and variance
\bea
\lav \stud(x) \rav_{\stud|D} \ \equiv \ \hat\stud(x|D) &=&
\kk(x)\T\KK^{-1}\yy 
\label{basic_inf_a}
\\
\lav (\stud(x)-\hat\stud(x))^2 \rav_{\stud|D} & = &
C(x,x) - \kk(x)\T\KK^{-1}\kk(x)
\label{basic_inf_b}
\eea
From the student's point of view, this solves the inference problem:
The best prediction for $\stud(x)$ on the basis of the data $D$ is
$\hat\stud(x|D)$, with a (squared) error bar given by\eq{basic_inf_b}.
The squared deviation between the prediction and the teacher is
$[\hat\stud(x|D)-\teach(x)]^2$; the generalization error (which, as a
function of $n$, defines the learning curve) is obtained by averaging
this over the posterior distribution of teachers, all datasets, and
the test input $x$:
\be
\eps = \lav \lav \lav [\hat\stud(x|D)-\teach(x)]^2
\rav_{\teach|D}\rav_D \rav_x
\label{eps_true}
\ee
Now of course the student does not know the true posterior of the
teacher; to estimate $\eps$, she must assume that it is identical to
the student posterior, giving from\eq{basic_inf_b}
\be
\hat\eps = \lav \lav \lav
[\hat\stud(x|D)-\stud(x)]^2 \rav_{\stud|D} \rav_D \rav_x = 
\lav \lav C(x,x) - \kk(x)\T\KK^{-1}\kk(x) \rav\XX \rav_x
\ee
where in the last expression I have replaced the average over $D$ by
one over the training inputs since the outputs no longer appear. If
the student model matches the true teacher model, $\eps$ and
$\hat\eps$ coincide and give the Bayes error, \ie\ the best achievable
(average) generalization performance for the given teacher.

I assume in what follows that the teacher is also a GP, but with a
possibly different covariance function $C_*(x,x')$ and noise level
$\noisest$. This allows eq.\eq{eps_true} for $\eps$ to be simplified,
since by exact analogy with the argument for the student posterior
\[
\lav \teach(x) \rav_{\teach|D}   \!=\! \kk_*(x)\T\KK_*^{-1}\yy, \quad
\lav \teach^2(x) \rav_{\teach|D} \!=\! \lav \teach(x) \rav_{\teach|D}^2 +
C_*(x,x) - \kk_*(x)\T\KK_*^{-1}\kk_*(x)
\]
and thus, abbreviating ${\bf a}(x) = \KK^{-1}\kk(x)-\KK_*^{-1}\kk_*(x)$,
\[
\eps = \lav \lav 
%
%
{\bf a}(x)\T\yy\yy\T{\bf a}(x) +
C_*(x,x) - \kk_*(x)\T\KK_*^{-1}\kk_*(x)
\rav_D \rav_x
\]
Conditional on the training inputs, the training outputs have
a Gaussian distribution given by the true (teacher) model; hence
$\lav\yy\yy\T\rav_{\{\yl\}|\{\xl\}} = \KK_*$, giving
\be
\eps = \lav \lav
C_*(x,x) - 2 \kk_*(x)\T\KK^{-1}\kk(x) + \kk(x)\T\KK^{-1}\KK_*\KK^{-1}\kk(x)
\rav\XX \rav_x
\label{eps_simp}
\ee

\section{Calculating the learning curves}

An exact calculation of the learning curve $\eps(n)$ is difficult
because of the joint average in\eq{eps_simp} over the training inputs
$X$ and the test input $x$. A more convenient starting point is
obtained if (using Mercer's theorem)
we decompose the covariance function into its eigenfunctions
$\phi_i(x)$ and eigenvalues $\Lam_i$, defined w.r.t.\ the input
distribution so that $\lav C(x,x')\phi_i(x')\rav_{x'} = \Lam_i
\phi_i(x)$ with the corresponding normalization $\lav
\phi_i(x)\phi_j(x) \rav_x=\delta_{ij}$. Then
\be
C(x,x')= \sum_{i=1}^\infty \Lam_i \phi_i(x) \phi_i(x'), \quad
\mbox{and similarly}\ 
C_*(x,x')= \sum_{i=1}^\infty \Lam^*_{i} \phi_i(x) \phi_i(x')
\label{c_decomp}
\ee
For simplicity I assume here that the student and teacher covariance
functions have the {\em same} eigenfunctions (but different
eigenvalues). This is not as restrictive as it may seem; several
examples are given below. The averages over the test input $x$
in\eq{eps_simp} are now easily carried out: E.g.\ for the last term we
need
\[
\lav (\kk(x)\kk(x)\T)_{lm} \rav_x = \sum_{ij} \Lam_i \Lam_j 
\phi_i(\xl) \lav\phi_i(x)\phi_j(x)\rav_x \phi_j(\xm) = 
\sum_i \Lam_i^2 \phi_i(\xl) \phi_i(\xm) 
\]
Introducing the diagonal eigenvalue matrix $(\LL)_{ij}=\Lam_i
\delta_{ij}$ and the `design matrix' $(\Ph)_{li}=\phi_i(\xl)$, this
reads $\lav \kk(x)\kk(x)\T \rav_x = \Ph\LL^2\Ph\T$. Similarly, for the
second term in\eq{eps_simp}, $\lav \kk(x)\kk_*\T(x) \rav_x =
\Ph\LL\LL_*\Ph\T$, and $\lav C_*(x,x) \rav_x = \tr \LL_*$. This gives,
dropping the training inputs subscript from the remaining average,
\[
\eps = \lav
\tr\LL_* - 2\, \tr \Ph\LL\LL_*\Ph\T\KK^{-1}
+ \tr \Ph\LL^2\Ph\T \KK^{-1} \KK_* \KK^{-1} \rav
\]
In this new representation we also have $\KK =
\noise\mident+\Ph\LL\Ph\T$ and similarly for $\KK_*$; for the inverse
of $\KK$ we can use the Woodbury formula
to write $\KK^{-1} = \noiseinv[\mident -
\noiseinv\Ph \Gfluc\Ph\T]$, where
$\Gfluc=(\LL^{-1}+\noiseinv\Ph\T\Ph)^{-1}$. Inserting these results,
one finds after some algebra that
\be
\eps = \noisest\noiseinv \left[\lav\tr\Gfluc\rav -
\lav\tr\Gfluc\LL^{-1}\Gfluc\rav\right] +
\lav\tr\Gfluc\LL_*\LL^{-2}\Gfluc \rav
\label{eps_transf}
\ee
which for the matched case reduces to the known result for the Bayes
error~\cite{Sollich99}
\be
\hat\eps = \lav\tr\Gfluc\rav
\label{eps_Bayes}
\ee
Eqs.\eqq{eps_transf}{eps_Bayes} are still exact. We now need to tackle
the remaining averages over training inputs. Two of these are of the
form $\lav \tr\Gfluc \MM \Gfluc\rav$; if we generalize the definition
of $\Gfluc$ to $\Gfluc=(\LL^{-1}+v\mident +w\MM+
\noiseinv\Ph\T\Ph)^{-1}$ and define $g=\lav\tr\Gfluc\rav$, then
they reduce to $\lav \tr\Gfluc \MM \Gfluc\rav=-\partial g/\partial
w$. (The derivative is taken at $v=w=0$; the idea behind introducing
$v$ will become clear shortly.) So it is sufficient to calculate $g$.
To do this, consider how $\Gfluc$ changes when a new example is added
to the training set. One has
\be
\Gfluc(n+1)-\Gfluc(n)=\left[\Gfluc^{-1}(n)+
\noiseinv\new\new\T\right]^{-1}-\Gfluc(n)=
-\ \frac
{\Gfluc(n) \new\new\T\Gfluc(n)} {\noise+\new\T\Gfluc(n)\new}
\label{G_update}
\ee
in terms of the vector $\new$ with elements $(\new)_i =
\phi_i(x_{n+1})$, using again the Woodbury formula. To obtain the
change in $g$ we need the average of\eq{G_update} over both the new
training input $x_{n+1}$ and all previous ones. This cannot be done
exactly, but we can approximate by averaging numerator and denominator
separately; taking the trace then gives $g(n+1)-g(n) =
-{\lav\tr\Gfluc^2(n)\rav}/[{\noise+g(n)}] $. Now, using our auxiliary
parameter $v$, $-\tr\lav\Gfluc^2\rav=\partial g/\partial v$; if we
also approximate $n$ as continuous, we get the simple PDE $\partial
g/\partial n = (\partial g/\partial v)/(\noise + g)$ with the initial
condition $g|_{n=0}=\tr(\LL^{-1}+v\mident +w\MM)^{-1}$. Solving this
using the method of characteristics~\cite{Sollich94f} gives a
self-consistent equation for $g$,
\be
{
g = \tr\left[\LL^{-1} + \left(v+\frac{n}{\noise+g}\right)\mident + w\MM\right]^{-1}
}
\label{g}
\ee
The Bayes error\eq{eps_Bayes} is $\hat\eps=g|_{v=w=0}$ and therefore
obeys
\be
{
\hat\eps = \tr\GG, \qquad \GG^{-1} = \LL^{-1} +
\frac{n}{\noise+\hat\eps}\,\mident
}
\label{hateps}
\ee
within our approximation (called `LC' in~\cite{Sollich99}). To obtain
$\eps$, we differentiate both sides of\eq{g} w.r.t.\ $w$, set $v=w=0$
and rearrange to give
\[
\lav \tr\Gfluc \MM \Gfluc\rav = -\partial g/\partial w =
(\tr\MM\GG^2)/[1-(\tr\GG^2)n/(\noise+\hat\eps)^2]
\]
Using this result in\eq{eps_transf}, with $\MM=\LL^{-1}$ and
$\MM=\LL^{-1}\LL_*\LL^{-1}$, we find after some further
simplifications the final (approximate) result for the learning curve:
\be
\eps = \hat\eps \,\frac
{\noisest \,\tr\GG^2 + n^{-1}(\noise+\hat\eps)^2 \,\tr\LL_*\LL^{-2}\GG^2}
{\noise   \,\tr\GG^2 + n^{-1}(\noise+\hat\eps)^2 \,\tr\LL^{-1}\GG^2}
\label{final}
\ee
which transparently shows how in the matched case $\eps$ and
$\hat\eps$ become identical.

\section{Examples}

\begin{figure}
\begin{center}
\epsfig{file=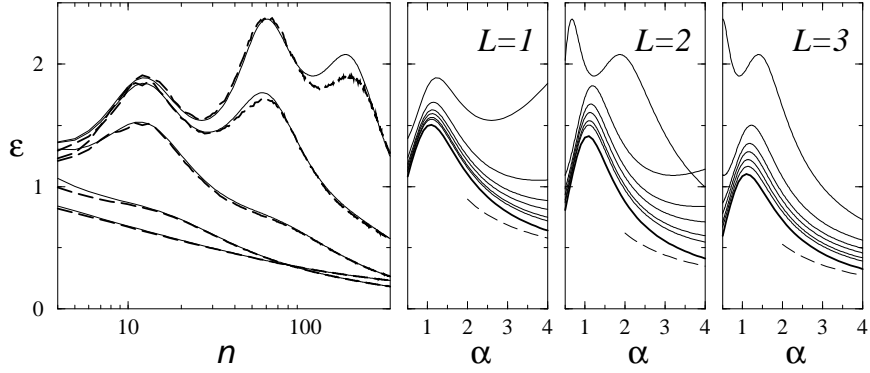, width=0.9\textwidth}
\end{center}
\vspace*{-0.8cm}
\caption{Left: Learning curves for RBF student and teacher, with
uniformly distributed, binary input vectors with $d=10$
components. Noise levels: Teacher $\noisest=1$, student
$\noise=10^{-4}, 10^{-3}, \ldots, 1$ (top to bottom). Length scales:
Teacher $l_* = {d}^{1/2}$, student $l=2{d}^{1/2}$. Dashed: numerical
simulations, solid: theoretical prediction. Right: Learning curves for
$\noise=10^{-4}$ and increasing $d$ (top to bottom: 10, 20, 30, 40,
60, 80, [bold] $\infty$). The $x$-axis shows $\alpha=n/\colvec{d}{L}$,
for learning stages $L=1,2,3$; the dashed lines are the universal
asymptotes~(\protect\ref{universal}) for $d\to\infty$.
\label{fig:binary}
}
\end{figure}

I now apply the result for the learning curve\eqq{hateps}{final} to
some exemplary learning scenarios. First, consider inputs $x$ which
are binary vectors%
\footnote{%
This scenario may seem strange, but simplifies the determination of
the eigenfunctions and eigenvalues; for large $d$, however, one
expects other distributions with continuously varying $x$ to give
similar results~\protect\cite{OppUrb01}.
} with $d$ components $x_a\in\{-1,1\}$, and assume that the input
distribution is uniform. We consider covariance functions for student
and teacher which depend on the product $x\cdot x'$ only; this
includes the standard choices (\eg\ OU and RBF) which depend on the
Euclidean distance $|x-x'|$, since $|x-x'|^2=2d-2x\cdot x'$. All these
have the same eigenfunctions~\cite{DieOppSom99}, so our above
assumption is satisfied. The eigenfunctions are indexed by subsets
$\rho$ of $\{1,2\ldots d\}$ and given explicitly by $\phi_\rho(x) =
\prod_{a\in\rho} x_a$. The corresponding eigenvalues depend only on
the size $\ii=|\rho|$ of the subsets and are therefore
$\colvec{d}{\ii}$-fold degenerate; letting $e=(1,1\ldots 1)$ be the
`all ones' input vector, they have the values $\Lam_\ii = \lav
C(x,e)\phi_\rho(x) \rav_x$ (which can easily be evaluated as an
average over two binomially distributed variables, counting the number
of $+1$'s in $x$ overall and among the $x_a$ with $a\in\rho$).
With the $\Lam_s$ and $\Lam^*_s$ determined, it is then a simple
matter to evaluate the predicted learning curve\eqq{hateps}{final}
numerically. First, though, focus on the limit of large $d$, where
much more can be said. If we write $C(x,x')=\cc(x\cdot x'/d)$, the
eigenvalues become, for $d\to\infty$, $\Lam_\ii = d^{-\ii}
\cc^{(\ii)}(0)$ and the contribution to $C(x,x)=\cc(1)$ from the
$\ii$-th eigenvalue block is $\lam_\ii\equiv\colvec{d}{\ii}\Lam_\ii
\to \cc^{(\ii)}(0)/\ii!$, consistent with $\cc(1) =
\sum_{\ii=0}^\infty \cc^{(\ii)}(0)/\ii!$ The $\Lam_\ii$, because of
their scaling with $d$, become infinitely separated for
$d\to\infty$. For training sets of size $n=\order(d^L)$, we then see
from\eq{hateps} that eigenvalues with $\ii>L$ contribute as if $n=0$,
since $\Lam_\ii\gg n/(\noise+\hat\eps)$; they have effectively not yet
been learned. On the other hand, eigenvalues with $\ii<L$ are
completely suppressed and have been learnt perfectly. We thus have a
hierarchical learning scenario, with different scalings of $n$ with
$d$---as defined by $L$---corresponding to different `learning
stages'. Formally, we can analyse the stages separately by letting
$d\to\infty$ at a constant ratio $\alpha=n/\colvec{d}{L}$ of the
number of examples to the number of parameters to be learned at stage
$L$ (note $\colvec{d}{L}=\order(d^L)$ for large $d$). An independent
(replica) calculation shows that our approximation for the learning
curve actually becomes {\em exact} in this limit. The resulting
$\alpha$-dependence of $\eps$ can be determined explicitly: Set $\cc_L
= \sum_{\ii\geq L} \lam_\ii$ (so that $\cc_0=\cc(1)$) and similarly
for $\cc^*_{L}$. Then for large $\alpha$,
\be
\eps = \cc^*_{L+1} + (\cc^*_{L+1}+\noisest)\alpha^{-1} +
\order(\alpha^{-2})
\label{universal}
\ee
This implies that, during successive learning stages, (teacher)
eigenvalues are learnt one by one and their contribution eliminated
from the generalization error, giving plateaux in the learning curve
at $\eps=\cc^*_1$, $\cc^*_2$, \ldots. These plateaux, as well as the
asymptotic decay\eq{universal} towards them, are
universal~\cite{OppUrb01}, \ie\ student-independent. The
(non-universal) behaviour for smaller $\alpha$ can also be fully
characterized: Consider first the simple case of linear perceptron
learning (see \eg~\cite{Sollich94f}), which corresponds to both
student and teacher having simple dot-product covariance functions
$C(x,x')=C_*(x,x')=x\cdot x'/d$. In this case there is only a single
learning stage (only $\lam_1=\lam^*_1=1$ are nonzero), and
$\eps=r(\alpha)$ decays from $r(0)=1$ to $r(\infty)=0$, with an
over-fitting maximum around $\alpha=1$ if $\noise$ is sufficiently
small compared to $\noisest$. In terms of this function $r(\alpha)$,
the learning curve at stage $L$ for {\em general} covariance functions
is then {\em exactly} given by $\eps = \cc^*_{L+1}+\lam^*_L r(\alpha)$
if in the evaluation of $r(\alpha)$ the effective noise levels
$\tilde\sigma^2 = (\cc_{L+1}+\noise)/\lam_L$ and $\tilde\sigma_*^2 =
(\cc^*_{L+1}+\noisest)/\lam^*_L$ are used. Note how in
$\tilde\sigma_*^2$, the contribution $\cc^*_{L+1}$ from the
not-yet-learned eigenvalues acts as effective noise, and is normalized
by the amount of `signal' $\lam^*_L=\cc^*_L-\cc^*_{L+1}$ available at
learning stage $L$. The analogous definition of $\tilde\sigma^2$
implies that, for small $\noise$ and depending on the choice of
student covariance function, there can be arbitrarily many learning
stages $L$ where $\tilde\sigma^2\ll\tilde\sigma_*^2$, and therefore
{\em arbitrarily many over-fitting maxima} in the resulting learning
curves. Fig.~\ref{fig:binary}(left) demonstrates that this conclusion
holds not just for $d\to\infty$; even for the cases shown, with
$d=10$, up to three over-fitting maxima are apparent. Our theory
provides a very good description of the numerically simulated learning
curves even though, at such small $d$, the predictions are still
significantly different from those for $d\to\infty$ (see
Fig.~\ref{fig:binary}(right)) and therefore not guaranteed to be
exact.

In the second example scenario, I consider continuous-valued input
vectors, uniformly distributed over the unit interval $[0,1]$;
generalization to $d$ dimensions ($x\in[0,1]^d$) is straightforward.
For covariance functions which are stationary, \ie\ dependent on $x$ and
$x'$ only through $x-x'$, and assuming periodic boundary conditions
(see~\cite{Sollich99} for details), one then again has covariance
function-independent eigenfunctions. They are indexed by integers%
\footnote{%
Since $\Lam_q=\Lam_{-q}$, one can assume $q\geq 0$ if all $\Lam_q$ for
$q>0$ are taken as doubly degenerate.
}
$q$, with $\phi_q(x) = e^{2\pi i q\cdot x}$; the corresponding
eigenvalues are $\Lam_q = \int \!dx \, C(0,x) e^{-2\pi i q\cdot x}$.
For the (`periodified') RBF covariance function $C(x,x') =
\exp[-(x-x')^2/(2l^2)]$, for example, one has $\Lam_q \propto
\exp(-\tilde q^2/2)$, where $\tilde q = 2\pi l q$. The OU case
$C(x,x')=\exp(-|x-x'|/l)$, on the other hand, gives $\Lam_q \propto
(1+\tilde q^2)^{-1}$, thus $\Lam_q \propto q^{-2}$ for large $q$.  I
also consider below covariance functions which interpolate in
smoothness between the OU and RBF limits: E.g.\ the MB2 (modified
Bessel) covariance $C(x,x') = e^{-a}(1+a)$, with $a=|x-x'|/l$, yields
functions which are once differentiable~\cite{WilViv00}; its
eigenvalues $\Lam_q \propto (1+\tilde q^2)^{-2}$ show a faster
asymptotic power law decay, $\Lam_q \propto q^{-4}$. To subsume all
these cases I assume in the following analysis of the general shape of
the learning curves that $\Lam_q \propto q^{-r}$ (and similarly
$\Lam^*_q \propto q^{-r_*}$). Here $r=2$ for OU, $r=4$ for MB2, and
(due to its faster-than-power law decay of eigenvalues) effectively
$r=\infty$ for RBF.

\begin{figure}
\begin{center}
\epsfig{file=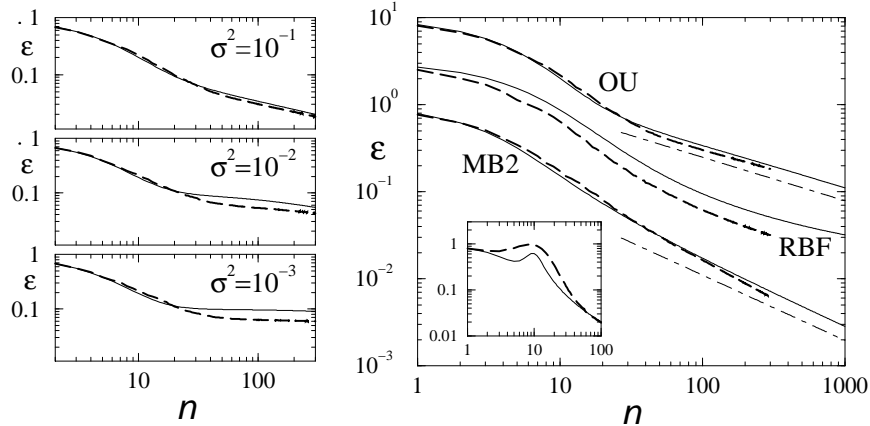, width=0.9\textwidth}
\end{center}
\vspace*{-0.8cm}
\caption{Learning curves for inputs $x$ uniformly distributed over
$[0,1]$. Teacher: MB2 covariance function, lengthscale $l_*=0.1$,
noise level $\noisest=0.1$; student lengthscale $l=0.1$ throughout.
Dashed: simulations, solid: theory. Left: OU student with $\noise$ as
shown. The predicted plateau appears as $\noise$ decreases. Right:
Students with $\noise=0.1$ and covariance function as shown; for
clarity, the RBF and OU results have been multiplied by $\sqrt{10}$
and $10$, respectively. Dash-dotted lines show the predicted
asymptotic power laws for MB2 and OU; the RBF data have a persistent
upward curvature consistent with the predicted logarithmic
decay. Inset: RBF student with $\noise=10^{-3}$, showing the
occurrence of over-fitting maxima.
\label{fig:MB2}
}
\end{figure}

From\eqq{hateps}{final}, it is clear that the $n$-dependence of the
Bayes error $\hat\eps$ has a strong effect on the true generalization
error $\eps$. From previous work~\cite{Sollich99}, we know that
$\hat\eps(n)$ has two regimes: For small $n$, where
$\hat\eps\gg\noise$, $\hat\eps$ is dominated by regions in input space
which are too far from the training examples to have significant
correlation with them, and one finds $\hat\eps\propto n^{-(r-1)}$. For
much larger $n$, learning is essentially against noise, and one has a
slower decay $\hat\eps \propto n^{-(r-1)/r}$. These power laws can be
derived from\eq{hateps} by approximating factors such as $[\Lam_q^{-1}
+ n/(\noise+\hat\eps)]^{-1}$ as equal to either $\Lam_q$ or to $0$,
depending on whether $n/(\noise+\hat\eps)<$ or $>\Lam_q^{-1}$. With
the same technique, one can estimate the behaviour of $\eps$
from\eq{final}. In the {\em small} $n$-regime, one finds $\eps \approx
c_1 \noisest + c_2 n^{-(r_*-1)}$, with prefactors $c_1$, $c_2$
depending on the student.  Note that the contribution proportional to
$\noisest$ is automatically negligible in the matched case (since then
$\eps = \hat\eps \gg \noise=\noisest$ for small $n$); if there is a
model mismatch, however, and if the small-$n$ regime extends far
enough, it will become significant. This is the case for small
$\noise$; indeed, for $\noise\to 0$, the `small $n$' criterion
$\hat\eps\gg\noise$ is satisfied for any $n$. Our theory thus predicts
the appearance of plateaux in the learning curves, becoming more
pronounced as $\noise$ decreases; Fig.~\ref{fig:MB2}(left) confirms
this%
\footnote{%
If $\noise=0$ exactly, the plateau will extend to $n\to\infty$. With
hindsight, this is clear: a GP with an infinite number of nonzero
eigenvalues has no limit on the number of its `degrees of freedom' and
can fit perfectly any amount of noisy training data, without ever
learning the true teacher function.}. Numerical evaluation also shows
that for small $\noise$, over-fitting maxima may occur before the
plateau is reached, consistent with simulations; see inset in
Fig.~\ref{fig:MB2}(right). In the {\em large} $n$-regime
($\hat\eps\ll\noise$), our theory predicts that the generalization
error decays as a power law. If the student assumes a rougher function
than the teacher provides ($r<r_*$), the asymptotic power law exponent
$\eps\propto n^{-(r-1)/r}$ is determined by the student alone. In the
converse case, the asymptotic decay is $\eps\propto n^{-(r_*-1)/r}$
and can be very slow, actually becoming logarithmic for an RBF student
($r\to\infty$). For $r=r_*$, the fastest decay for given $r_*$ is
obtained, as expected from the properties of the Bayes error. The
simulation data in Fig.~\ref{fig:MB2} are compatible with these
predictions (though the simulations cover too small a range of $n$ to
allow exponents to be determined precisely).

In summary, the above approximate theory makes a number of non-trivial
predictions for GP learning with mismatched models, all borne out by
simulations: for large input space dimensions, the occurrence of
multiple over-fitting maxima; in lower dimensions, the generic
presence of plateaux in the learning curve if the student assumes too
small a noise level $\noise$, and strong effects of model mismatch on
the asymptotic learning curve decay. The behaviour is much richer than
for the matched case, and could guide the choice of (student) priors
in real-world applications of GP regression; RBF students, for
example, run the risk of very slow (logarithmic) decay of the learning
curve if the target (teacher) is less smooth than assumed. An
important issue for future work is to analyse to which extent
hyperparameter tuning (\eg\ via evidence maximization) can make GP
learning robust against some forms of model mismatch, \eg\ a
misspecified functional form of the covariance function.

\vspace*{-2\baselineskip}
\small

\begin{thebibliography}{10}

\bibitem{general_refs}
See \eg\ D J C MacKay, Gaussian Processes, Tutorial at {\em NIPS 10}; recent
  papers by Malzahn/Opper (in {\em NIPS 13}), Csat\'{o} {\it et al.} ({\em NIPS
  12}), Goldberg/Williams/Bishop ({\em NIPS 10}), Williams and
  Barber/Williams ({\em NIPS 9}), Williams/Rasmussen ({\em NIPS 8}); and
  references below.
%
\bibskip
%
\bibitem{MicWah81}
C~A Michelli and G~Wahba.
\newblock In Z~Ziegler, editor, {\em Approximation theory and applications},
  pages 329--348. Academic Press, 1981;
M~Opper. 
\newblock In I~K Kwok-Yee {\it et al.}, editors, {\em
  Theoretical Aspects of Neural Computation},
  pages 17--23. Springer, 1997.
%
\bibskip
%
\bibitem{Sollich99}
P~Sollich.
In {\em NIPS 11}, pages 344--350.
%
\bibskip
%
\bibitem{WilViv00}
C~K~I Williams and F~Vivarelli.
\newblock {\em Mach.\ Learn.}, 40:77--102, 2000.
%
\bibskip
%
\bibitem{Williams98}
C~K~I Williams.
\newblock In M~I Jordan, editor, {\em Learning and Inference in Graphical
  Models}, pages 599--621. Kluwer Academic, 1998.
%
\bibskip
%
%
\bibitem{Sollich94f}
P~Sollich.
\newblock {\em J.\ Phys.\ A}, 27:7771--7784, 1994.
%
\bibskip
%
\bibitem{OppUrb01}
M~Opper and R~Urbanczik.
\newblock {\em Phys.\ Rev.\ Lett.}, 86:4410--4413, 2001.
%
\bibskip
%
\bibitem{DieOppSom99}
R~Dietrich, M~Opper, and H~Sompolinsky.
\newblock {\em Phys.\ Rev.\ Lett.}, 82:2975--2978, 1999.

\end{thebibliography}

\newcommand{\bibskip}{\vspace*{-0.1\baselineskip}}

\end{document}